\documentstyle[12pt,epsf]{article}

\def\Ex#1{\langle#1\rangle}
\def\qed{\nobreak\kern 1em \vrule height .5em width .5em depth 0em}
\def\vbar{\mathchoice{\vrule height6.3ptdepth-.5ptwidth.8pt\kern-.8pt}
   {\vrule height6.3ptdepth-.5ptwidth.8pt\kern-.8pt}
   {\vrule height4.1ptdepth-.35ptwidth.6pt\kern-.6pt}
   {\vrule height3.1ptdepth-.25ptwidth.5pt\kern-.5pt}}

\def\date
   {\noindent Date: \today \par
    \medskip}

\newcount\subno      
\newcount\subsubno   
\newcount\appno      
\newcount\appeqno    
\newcount\tableno    
\newcount\figureno   
\begin{document}

\begin{center}
{\large \bf Inhomogeneous Reptation of Polymers} \\[6mm]
M.J.E. Richardson\\
{\small \em Theoretical Physics, University of Oxford,\\
1 Keble Road, Oxford, OX1 3NP, United Kingdom}\\[4mm]
and\\[4mm]
G.M. Sch\"utz\\
{\small \em IFF, Forschungszentrum J\"ulich, 52425 J\"ulich, Germany}
\end{center}

\begin{abstract}
We study the motion of long polymers (eg DNA) in a gel under the influence
of an external force acting locally on small segments of the
polymer. In particular, we examine the dependence of the drift
velocity on the position where the force acts and the length of the
polymer. As an application, we discuss the possibility of {\em
gel magnetophoresis} - the size-separation of
long polymers by the attachment of a magnetic bead at an 
arbitrary position along the length of the polymer. We show that there is a regime where the separation 
of such polymers with this `random-position beading' is possible. 
\end{abstract}


\section{Introduction}

Gel electrophoresis is a widely used technique for the
separation of long chain polymers such as DNA. This provides a
motivation for the understanding of the microscopic physics involved
in the motion of polymers in gels and similar structures under the
influence of an external driving force. The notions of {\it reptation} and
the {\it confining tube} have emerged as fruitful concepts
\cite{deGe71,Edwa67}. However, as shall be discussed in this paper, if a
driving force only acts locally on small polymer segments rather than
homogeneously along its whole length, one has to include the formation
of {\it hernias} into the picture. This leads to new and interesting
behaviour for the drift velocity of the polymers under the action of
such a force. 

Gel electrophoresis works by sifting the
polymers through a gel in the presence of an electric field. Polymers
of different lengths travel at different velocities, allowing the
resolution of an initial multi-length mixture. However, if the
polymers are too long, it becomes impossible to separate them. The
velocity becomes length independent - a phenomenon known as band collapse.
There are various techniques available to overcome band collapse, for
example pulsed field techniques. However, recently it was noted that 
this could be done in a much simpler way by attaching a 
magnetic bead to the end of a polymer and
pulling on the bead with a magnetic field. In this way band collapse can be
avoided \cite{Bark96b}. Hence, one could separate DNA of large lengths
using the technique where a force acts only on one small segment
of the polymer.

It may be difficult to attach beads specifically to the end of a polymer 
and therefore it is worth studying the effect
of a magnetic field if a magnetic bead is attached at an arbitrary position 
along the polymer chain. In this case one would expect the drift velocity
resulting from the force on the bead to have some dependence on its position.
This effect of {\em random-position beading} magnetophoresis as opposed to the 
chain-end beading of Ref. \cite{Bark96b} is what we wish to study in this paper.
For reasons discussed below we have to consider only the situation where a
single bead is attached to the DNA. We would like to stress that in other
applications of tags in electrophoresis the force acts on the polymer 
\cite{Ulan90,Nool92,Maye94,Adja95,Aalb95} but not on the tag as proposed in 
\cite{Bark96b} and here.

If it were necessary to separate a mixture of two long polymers of different 
lengths one needs to consider the relative sizes of the dispersion in the 
velocity $\Delta{V(k)}_{N}$ due to the spread of different positions $k$ of 
the bead for a polymer of given length $N$ and the length dependent velocity 
$V(N)$. If the random-position beading is to be useful it must fulfill the 
following criteria for viability:
\begin{itemize}
\item The polymers must move through the gel at a reasonable speed.
\item For a given set of polymer lengths $N_{i}$, the
dispersion $\Delta{V(k)}_{N_{i}}$ due to the different positions of the 
beads for a given $N_{i}$ should not overlap with the velocity $V_{N_{i}}$ for 
polymers of different length.
\end{itemize}
If the second condition is not fulfilled bands of polymers of different
length will overlap.
In the following work the behaviour of the velocities as a function of
$k$ and $N$ will be examined and it will be shown that there is a
regime where this type of beading allows separation. We will focus on the
application to DNA, but this kind of magnetophoresis may be of much broader
relevance.

\section{The Microscopic Physics and the Model}

For a theoretical study of magnetic separation one needs a model which 
provides an adequate {\em quantitative} description of the motion of 
DNA in a gel matrix. As a starting point we shall use the Rubinstein-Duke
model \cite{Rubi87,Duke89} for gel electrophoresis. This model has been studied
extensively both analytically \cite{Rubi87,Wido91,vanL91,vanL92,Prae94,Leeg95,Wido96} and 
numerically \cite{Duke89,Wido91,Bark94}. Predictions for the drift velocity of 
a polymer in a constant electric field obtained in Monte Carlo simulations from 
this model \cite{Bark94} have been shown to be in excellent agreement 
with experimental data \cite{Bark96a}. This makes this model a 
candidate for suitably chosen generalizations.

The essential ingredients are the notion of a confining tube and motion by 
reptation, concepts introduced by Edwards \cite{Edwa67} and
de Gennes \cite{deGe71}. The DNA itself is divided into units of its 
persistence length, called reptons, and is thus viewed as a chain of $N$ reptons
moving wormlike through the pores of a rigid network of entangled polymers 
(the gel strands). In this
simple picture the DNA is confined to a tube, the shape of which is defined
by the actual spatial configuration of the DNA in the gel. It is assumed that
only the ends of the tube - the head and tail of the DNA - can move and
change the shape of the tube. For example, one end of the DNA may
recede into the interior of the existing tube and then move forward
again into a different direction thus changing the configuration. The interior of the DNA is only allowed to move along the
existing tube. Reptons can hop between neighbouring pores within the tube, 
but not sideways as a lateral excursion would involve breaking out of the tube. 
Such creation of {\em hernias} is certainly
taking place in a real system, but is neglected within this model.

Rubinstein \cite{Rubi87} introduced a further simplification by depicting the
random gel network as a regular cubic lattice in which the unit cells are the 
pores and the links are the gel strands (see Fig.~1). This lattice model was 
further developed by Duke \cite{Duke89} who introduced local detailed balance 
for the stochastic motion of the reptons between pores in the presence of an 
electric field. Since DNA is an acid, it carries a negative electric charge in 
a suitably chosen buffer solution. This overall charge is divided equally on
each repton which is thus driven against an external electric field.
The rules according to which the polymer moves may be summarised as follows:

\noindent
(a) Reptons in the interior of the chain move only along the sequence
of pores occupied by the chain. This restriction ensures that the only
possible mechanism of movement is {\it reptation}.

\noindent
(b) At least one repton must remain in each pore along the chain but
otherwise the number of reptons in a pore is unrestricted. By imposing
this rule we achieve that the polymer has some elasticity, but does not
stretch to infinite length.

\noindent
(c) The two chain end segments may move to adjacent pores provided that
the rule (b) is not violated.

Each possible move
(each repton moving in any direction) is tried stochastically with a constant rate,
setting the unit of time. This represents the diffusive motion of the
polymer. If a force acts on the reptons, the rate for hopping in direction
of the force is larger than the unit rate, while the hopping rate against
the force becomes smaller. 

It may be worthwhile pointing out that in this model very many reptons or even 
all of them, ie the whole polymer, may be found within a single pore (see rule
(b)) which is clearly unphysical. However, for a long polymer consisting
of many reptons such an event would be extremely rare under the dynamics
described above. Consequently, its contribution to the motion of the
polymer is insignificant. Such a situation 
could in fact be completely suppressed by introducing a force preventing
occupancy of the same pore by more than a certain maximal number of reptons.
Considering the accuracy of the model as seen in Refs. \cite{Bark96b,Bark96a}
such a modification might perhaps improve the model slightly but does not seem 
to be an urgent necessity.

While this simple picture of the confining tube appears to be adequate for the
quantitative description of ordinary constant-field electrophoresis, this
can clearly not be the case if a force acts only on one or very few reptons in the bulk of the polymer. One would intuitively expect that in such a 
situation the formation of 
hernias by these special reptons will dominate the overall motion of the DNA.
This is indeed the situation that we propose to study. Attaching
a magnetic bead at some position along the DNA and exerting a force with a
magnetic field corresponds to pulling on a single repton in the framework
of the model discussed above. Therefore, we have to modify the original
Rubinstein-Duke model in a suitable way. We shall choose the simplest possible
modification, that is we assume that no force acts on reptons except
on a single
tagged one. We still assume that no hernias are formed except by this tagged
repton. For further simplification we shall assume that this repton shall
never hop against the force direction. This means that a hernia, once formed,
will not disappear. In an experimental setup this may be approximately achieved
by exerting a sufficiently strong force combined with a suitably
chosen mobility for the
beaded repton. We do not expect this to be an oversimplification, since
back hopping of the tagged repton will presumably change only the overall time
scale of the motion, and not the qualitative features of this situation.

Our generalization may thus be summarized as follows (see also Fig.~1):
\begin{itemize}
\item Only the case of one magnetic bead attached to the polymer is 
considered. This will be justified in section 5.
\item Only the magnetic bead experiences a force.
\item It is only allowed to move in the direction of the force.
\item It is able to break through the confining tube. However, as
formation of hernias is unfavourable the rate at which this occurs
will be less than that for diffusion of reptons in the tube.
\item The normal rules of the zero-field repton model apply to all
reptons that are not directly attached to the magnetic bead.
\end{itemize}

Since we are only interested in the motion of the DNA in the direction of the
force (which we define to be the $x$-direction) we may use the standard mapping 
of the $x$-coordinates $x_k$ of the $N$ reptons onto a hard core diffusion
process on a one-dimensional lattice of $L=N-1$ sites, see e.g. \cite{Bark94}, 
and generalize it to allow for the motion of the tagged repton out of the tube.
We shall assume the bead to be attached to repton $k+1$ where because of 
reflection symmetry we assume $1 \leq k \leq L/2$.
In this mapping where an $A$
denotes a particle and a $0$ denotes a hole, we have in the bulk
$$
A0 \leftrightarrow 0A \hspace{10mm} \mbox{with rate $1$,}
$$
except at bond $k,k+1$ where
$$
00 \rightarrow AA \hspace{10mm} \mbox{with rate $r$.}
$$
The creation of two particles corresponds to breaking out of the tube to form
a hernia as can be seen in Fig.~1. The rate $r$ is phenomenological and has to
be fitted to experimental data. 
At the boundaries particles are allowed to diffuse in and out of
the confining tube
$$
\begin{array}{ccc}
0 & \rightarrow & A \hspace{10mm} \mbox{with rate $\alpha$,}\\
A & \rightarrow & 0 \hspace{10mm} \mbox{with rate $\beta$.}
\end{array} 
$$

This mapping has the advantage of reducing the calculation of the
drift velocity in a three-dimensional model
to the analysis of a one-dimensional lattice gas problem which can be studied
numerically and analytically. In this mapping the average drift
velocity $V$
is given in units of the lattice by the steady state current of
particles $C$ induced by the creation
of particle pairs at sites $k,k+1$. 

In deriving these rules we have assumed that the tagged repton is always
ahead of or at least in the same pore as its neighboring reptons. The reason for
this is that once the polymer is in such a configuration
it can escape from it only if {\em all} reptons $\{1,\dots,k\}$ to the left
or {\em all reptons} $\{k+2,\dots, N\}$ to the right of the
tagged repton move into the same pore as the tagged one. Unless the tagged
repton is very close to one end of the polymer this cannot happen in the
real system and, as discussed above, is exceedingly unlikely to happen within the 
framework of our model. This justifies the simplification made here.
On the other hand, if the tagged repton is close
to the end of the polymer, the results of Ref. \cite{Bark96b} apply which we
recover in the limit $k$ small and $L$ large.
Note that in a general consideration of hernia formation one must keep
a count of each hernia's position. This involves pairing particles
$\tilde{A},\tilde{B}$ on either side of each hernia. However, under the
assumptions made above, ie allowing hernia production only at the
beaded repton and also only allowing this beaded repton to move
forward, the complication of pairing particles becomes unnecessary. 

\newpage
\begin{figure}[h]
\epsfxsize 13 cm
\epsfysize 10.5 cm
\epsfbox{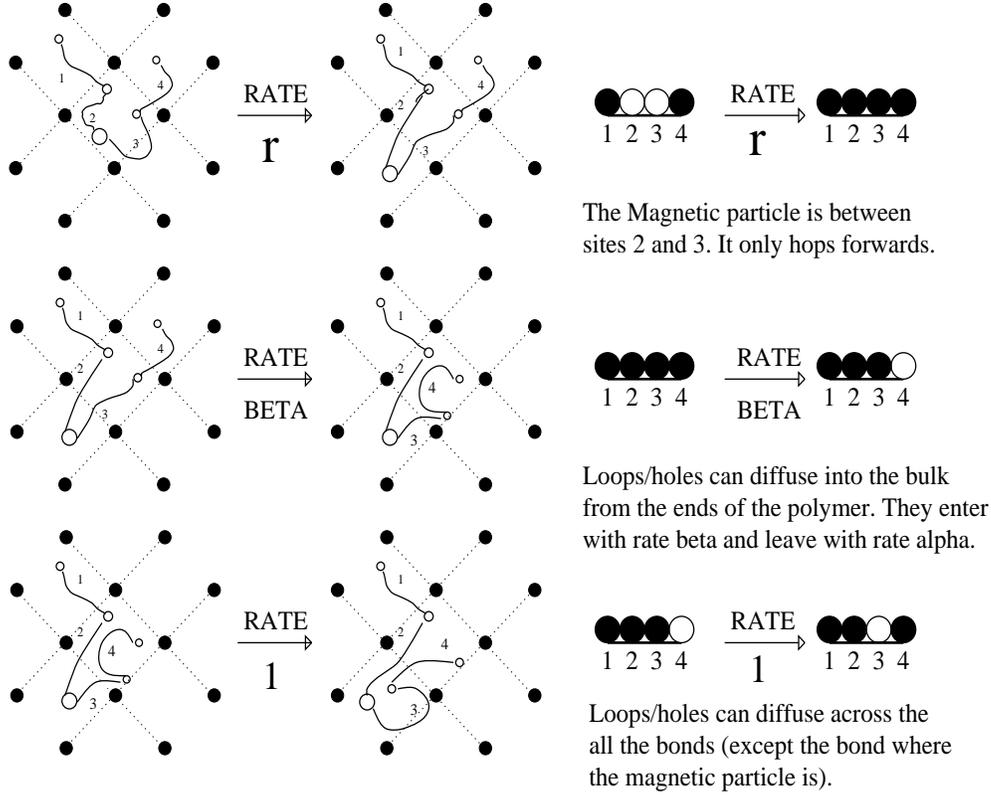}
\caption{The allowed moves showing the dynamics of a short polymer with a single magnetic bead attached
on repton 3, ie between sites 2 and 3. Hernias (excursions out of
the confining tube) occur only at the position of the magnetic bead as
this is the only part that experiences a force. The velocity of the polymer as
a whole is proportional to the density of loops of stored length at
the position of the bead.}
\end{figure}
\section{The Currents}

Let $n_{i}$ be the binary random variable denoting an occupation of site $i$
by a particle, ie $n_i = 1$ if site $i$ is occupied and $n_i=0$ if it is
vacant. For convenience we also introduce $v_i = 1-n_i$. The process described 
above leads to the following exact evolution equations for the particle density
\begin{eqnarray}
\frac{\partial{\Ex{n_{1}}}}{\partial t} & = & {\alpha}\Ex{v_{1}} - 
{\beta}\Ex{n_{1}} +
\Ex{n_{2}} - \Ex{n_{1}} \\
\frac{\partial{\Ex{n_{i}}}}{\partial t} & = & \Ex{n_{i-1}} - \Ex{n_{i}} + 
\Ex{n_{i+1}} - \Ex{n_{i}} \\
\frac{\partial{\Ex{n_{k}}}}{\partial t} & = & \Ex{n_{k-1}} - \Ex{n_{k}}
+r\Ex{v_{k}v_{k+1}} \\
\frac{\partial{\Ex{n_{k+1}}}}{\partial t} & = & \Ex{n_{k+1}} - \Ex{n_{k}}
+r\Ex{v_{k}v_{k+1}} \\
\frac{\partial{\Ex{n_{L}}}}{\partial t} & = & {\alpha}\Ex{v_{L}} - 
{\beta}\Ex{n_{L}} +
\Ex{n_{L-1}} - \Ex{n_{L}}
\end{eqnarray}
This is a non-equilibrium system with a constant steady state current $C$ of
particles.
As the currents flow from the bond $k,k+1$ it is useful to define 
$C$, which is proportional to the velocity $V$ of the polymer, as
\begin{eqnarray}
C  & = & {\beta}\Ex{n_{1}} - {\alpha}\Ex{v_{1}} \\
C  & = & \Ex{n_{i+1}} - \Ex{n_{i}}, \mbox{ for } i<k \\ 
C  & = & r\Ex{v_{k}v_{k+1}}  \\
C  & = & \Ex{n_{j}} - \Ex{n_{j+1}}, \mbox{ for } j>k+1\\
C  & = & {\beta}\Ex{n_{L}} - {\alpha}\Ex{v_{L}}
\end{eqnarray}
Ignoring correlations at the $k,k+1$ bond - ie approximating
$\Ex{v_{k}v_{k+1}}=\Ex{v_{k}}\Ex{v_{k+1}}$ we obtain
a simple closed recursion for the steady state particle density and find
\begin{eqnarray}
\Ex{n_{1}} &=&  \frac{C}{\alpha + \beta} + \frac{\alpha}{\alpha + \beta} \\	
\Ex{n_{L}} &=& \Ex{n_{1}} \\	
\Ex{n_{k}} &=& (k-1)C + \Ex{n_{1}} \\
\Ex{n_{k+1}} &=& (L-k+1)C + \Ex{n_{L}} \\
\frac{C}{r} &=& (1-\Ex{n_{k}})(1-\Ex{n_{k+1}})
\end{eqnarray}
\newpage
\begin{figure}[t]
\epsfxsize 11 cm
\epsfysize 7.5 cm
\epsfbox{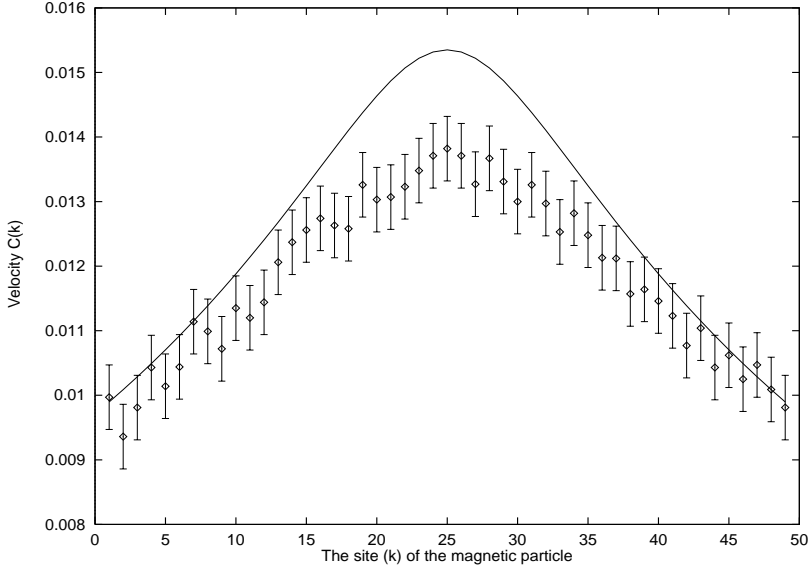}
\caption{Velocity $C$ in lattice units as a
function of the position of the attached magnetic bead ($k$) for
$L=50$, $r=1$ and $\alpha=\beta=1$ given by mean-field (Eq. 
(\protect\ref{C})) and
Monte Carlo simulations. The feature seen here of $C_{mf}{\ge}C_{MC}$ is 
generic for all values of $r$, the equality
holding when the bead is attached at either end \protect\cite{Bark96b}. Hence,
the mean field solution gives greater dispersion than the exact model
and can be used as a bound.}
\end{figure}

Defining $a=\frac{\beta}{\alpha + \beta}$ and $b=\frac{1}{\alpha +
\beta}-1$ we get for the current
\begin{eqnarray}
C &=& \frac{a(L+2b) + r^{-1}}{2(k+b)(L-k+b)} \nonumber \\
\label{C}
&-& \frac{\sqrt{(a(L+2b) + r^{-1})^{2} - 4a^{2}(k+b)(L-k+b)}}{2(k+b)(L-k+b)}
\end{eqnarray}
This gives the drift velocity of the DNA with a magnetic bead attached to
repton $k+1$ as a function of the position $k$, the length $N=L+1$ of the
DNA and the effective hopping rate $r$ of the tagged repton, see Fig.~2. Not surprisingly the current reaches a maximum if the
bead is attached to the center of the DNA and decreases monotonically
with increasing distance from the center. In the example given in Fig.~2
the current for the system with a bead attached to the end of the DNA
as studied in Ref. \cite{Bark96b} is approximately two thirds of the current
of the same DNA with a bead attached at the center. As one can see from Monte 
Carlo simulations, the mean field approximation made above by ignoring the
correlation between sites $k,k+1$ leads to a current which is slightly higher
than the true current and also enhances the difference in current between
chain-end beading and central beading. Therefore, the exact difference between
these two extremal situations is actually smaller than given by 
the expression (\ref{C}) and the mean-field approximation can be used
as a bound on the exact model.

\section{Dispersion and Separability}

In the following analysis we can for simplicity set $b=0$ corresponding to $\alpha +\beta=1$. This is just a statement 
about the relative timescales between the boundary and the bulk and
will not effect the general type of behaviour.\footnote{Choosing $\alpha=2$ and
$\beta=1$ corresponds to the choice of rates in
Ref. \protect\cite{Bark96b}. In each of the figures we have chosen $\alpha=\beta=1$.} For consistency of notation we
express the length dependence of the velocity in terms of $L$ rather than
in terms of the number of reptons $N=L+1$.
The greatest dispersion for a given $k$ as $L$ is kept constant is
between the cases where $k=1$ (at an end) and $k=\frac{L}{2}$ (in the
middle). Denoting the current for these two cases by $C_{e}$ and $C_{m}$ 
respectively we have (with $b=0$)
\begin{eqnarray}
C_{e} &=& \frac{aL + r^{-1} - \sqrt{(aL + r^{-1})^{2} - 4a^{2}(L-1)}}{2(L-1)}
\end{eqnarray}
\begin{eqnarray}
C_{m} &=& \frac{aL + r^{-1} - \sqrt{(aL + r^{-1})^{2} - 
(aL)^{2}}}{2(\frac{L}{2})^{2}}
\end{eqnarray}

If we want to separate polymers of length $L_{1}$ and $L_{2}$ it is
important that the dispersion due to differing positions ${\Delta}k$ for a given
$L$ of the bead is less than their separability due to their different lengths
${\Delta}L$. Some examples of $C_{e}(L)$ and $C_{m}(L)$ are given in 
Fig.~3. The velocities seen in this figure are of the same order as
those seen in gel electrophoresis, thus fulfilling the first criterion for
viability in section 1.
\newpage
\begin{figure}[t]
\epsfxsize 12  cm
\epsfysize 10 cm
\epsfbox{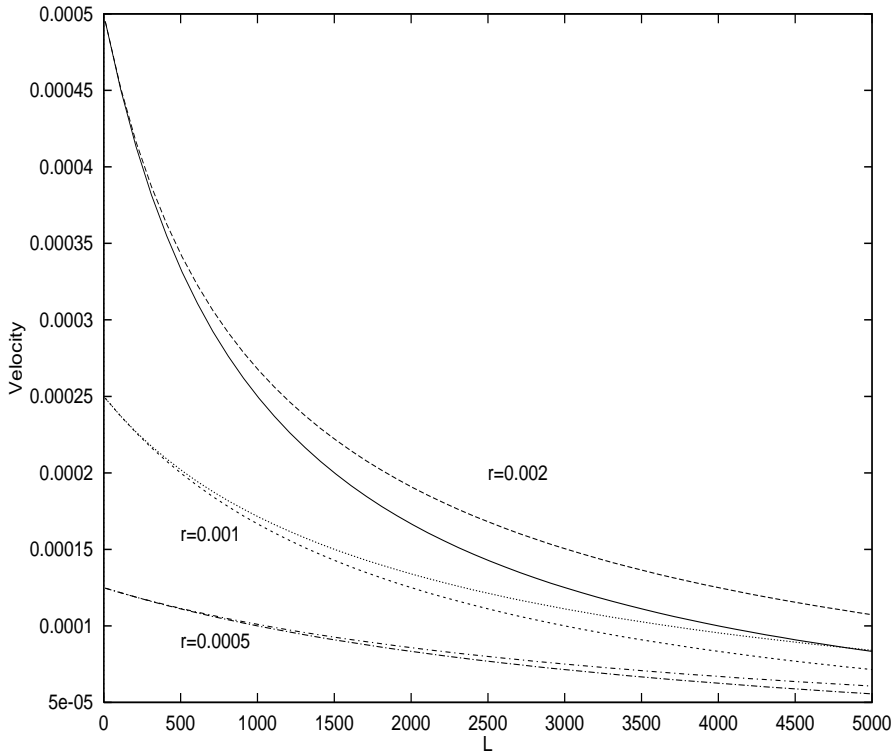}
\caption{Comparison of $C_{e}$ and $C_{m}$ for different $r$ as $L$
varies with $\alpha=\beta=1$. $C_{m}$ is always the greater in each pair. Good
separability occurs when the two lines in a pair $C_{e}$ and $C_{m}$
are close together as this represents low dispersion. Hence, one sees
a small $r$ separation regime. The velocities seen here are of the
same order as those in constant-field gel electrophoresis.}
\end{figure}
In order to study the separability due to this dispersion we recall that
$C_{e}(L)<C_{m}(L)$. Hence, if we are to separate polymers of length $L_{1}$ 
and $L_{2}$ where $L_{1}<L_{2}$ we require $\Delta{V} = C_{e}(L_{1}) - 
C_{m}(L_{2}) >0$. This inequality gives a lower bound for the length difference 
of two polymers for which separation is possible. 

For extremely small $r \ll (aL)^{-1}$, ie for very immobile beads,
one gets $C_{m} \approx C_{e} \approx a^2 r$. There is no length dispersion due
to the position of the bead, but the velocity is very small and length
independent. The motion of the DNA is completely determined by the motion
of the nearly immobile bead and one has band collapse. It is interesting
to note that this situation is analogous to that studied by Aalberts 
\cite{Aalb95} for an extremely bulky end group attached to the DNA under 
conditions encountered in standard electrophoresis where the force acts on the 
chain rather than on the bead.

For mobile beads $r \gg (aL)^{-1}$ one gets $L_2/L_1 > 2$, making separation
possible. Numerically, solutions of this inequality show that it is 
possible to satisfy the conditions given above also for $r$ of the order
$(aL)^{-1}$. In this case one can achieve smaller separation ratios of 
$L_2/L_1$. For a detailed numerical analysis, a wide range of $L_{1},L_{2}$ 
were used and the graphs all showed the same characteristics as seen in Fig.~4 
for $L_1=400$ and $L_2=500$ as a function of $1/r$. There is a regime where 
$\Delta{V}>0$ and, importantly,  also a particular value for $r$ which will 
maximise the separation. Thus, the second criterion for viablity in 
section 1 is also satisfied. 
\begin{figure}[h]
\epsfxsize 11.5 cm
\epsfysize 9.5 cm
\epsfbox{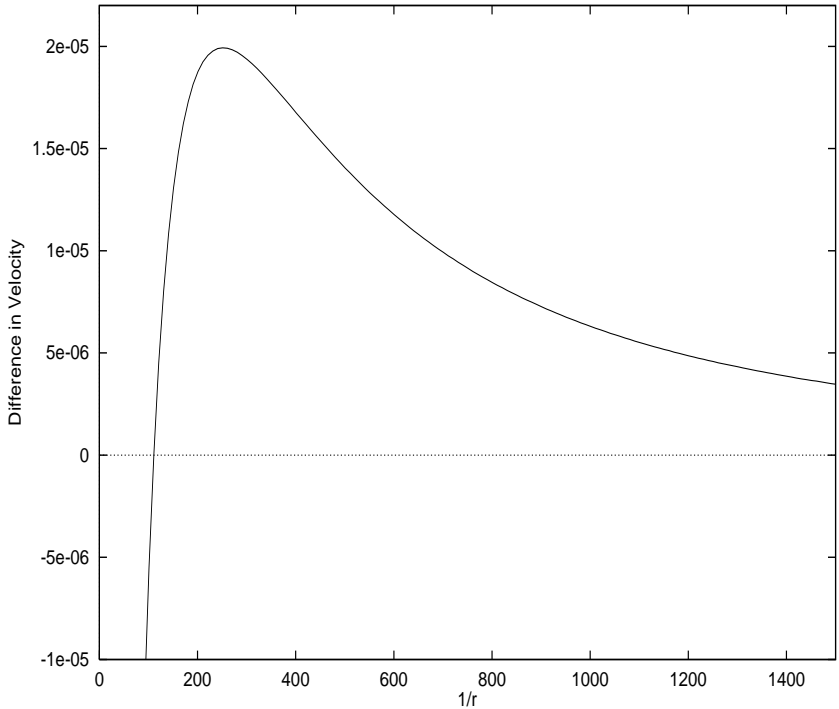}
\caption{Velocity difference for the fastest moving polymer of length
500 and the slowest moving polymer of length 400,
$C_{e}(400)-C_{m}(500)$ with $\alpha=\beta=1$. 
The two polymer fragments are separable for approximately $1/r \geq 100$. }
\end{figure}

\section{Conclusion}

To conclude, we have introduced a new variant of the Rubinstein-Duke model
to describe the motion of polymers in a gel matrix. We studied the situation
where a force acts locally on the polymer and drags it through the gel.
Unless one uses extremely immobile beads one finds that
even for strong forces there is no band collapse - the drift velocity
always depends on the length of the polymer. This is unlike standard (constant
field) gel electrophoresis where for sufficiently strong forces long polymers
travel with the same velocity and therefore do not separate.

Furthermore, we have shown that the drift velocity of the polymer depends not 
only on its length, but also on the position where the force acts. As a direct 
result of this, there is velocity dispersion in a mixture of polymers of the 
{\em same} length if the force acts on different positions in the individual 
polymers. Nevertheless, we have shown that there is still a regime for
which it is possible to separate polymers of different length by letting 
them move through the gel.

Experimentally, this situation could be realized by attaching magnetic beads
to DNA and using a magnetic field gradient \cite{Hauk93}. If a bead was made 
that would attach itself to a commonly occuring chemical group this could be 
mixed in with the DNA to produce polymers that were tagged at differing 
positions $k$ from the polymer ends. So for a given length $N$ of polymer there 
would be a spectrum of configurations with beads attached at different $k$. 
Some DNA fragments would have no beads, others would have one or more. Once 
placed in a gel with a magnetic field polymers with no beads will not experience
a force and polymers with 2 or more beads will become caught in the gel in
hook-like configurations. Such DNA fragments would not move at all. Hence, one 
only needs to consider the case of polymers that each have a single bead 
attached at some position $k$ - the situation studied in this paper.

\section*{Acknowledgments}
We would like to thank G. Barkema for useful
discussions. M.J.E.R. acknowledges financial support from the EPSRC
under Award No. 94304282.

\bibliographystyle{unsrt}

\end{document}